\documentclass[twocolumn,floatfix,preprintnumbers,superscriptaddress]{revtex4}

\usepackage[utf8]{inputenc}
\usepackage[colorlinks=true,citecolor=blue,linkcolor=blue]{hyperref}
\usepackage[normalem]{ulem}
\usepackage{url}
\usepackage{graphicx,wrapfig,float,slashed,cancel}
\usepackage{amsmath,amssymb,epsfig,graphicx,xcolor,stmaryrd}
\usepackage{bm}
\usepackage{enumitem}
\usepackage{xcolor}

\definecolor{darkblue}{RGB}{1, 90, 173}

\begin{document}


\title{Light quarkonium hybrid mesons}

\author{B. Barsbay\textsuperscript}
\affiliation{Division of Optometry, School of Medical Services and Techniques, Do\v{g}u%
\c{s} University, 34775 Istanbul, T\"{u}rkiye}

\author{K. Azizi} %
\thanks{Corresponding Author}
\affiliation{Department of Physics, University of Tehran, North Karegar Ave. Tehran 14395-547, Iran}
\affiliation{Department of Physics, Do\v{g}u\c{s} University, Dudullu-\"{U}mraniye, 34775
Istanbul, T\"{u}rkiye}
\affiliation{School of Particles and Accelerators, Institute for Research in Fundamental
Sciences (IPM) P.O. Box 19395-5531, Tehran, Iran}

\author{H.~Sundu}
\affiliation{Department of Physics Engineering, Istanbul Medeniyet University, 34700
Istanbul, T\"{u}rkiye}

\date{\today}

\preprint{}

\begin{abstract}
We investigate  the light quarkonium hybrid mesons of various spin-parities in QCD. Considering different interpolating currents made of the valence  light quarks and single gluon,  we calculate the mass and current coupling of the strange and nonstrange members of light hybrid mesons  by including into computations the nonperturbative  quark and gluon condensates  up to ten dimensions in order to increase the accuracy of the results. The obtained  results  may  be useful for future experimental searches of these hypothetical states.  They can also be used  in the calculations of  different parameters related to the decays/interactions of light hybrid mesons  to/with other states.
\end{abstract}


\maketitle

\section{Introduction} \label{sec:intro}

As  successful theory of strong interaction, the quantum chromodynamics (QCD) together with the powerful quark model have predicted the existence of exotic hadrons beyond the standard mesons and baryons. The most known categories for exotic hadrons are tetraquarks, pentaquarks, hexaquarks, quark-gluon hybrids and glueballs. Starting from 2003, many tetraquark and pentaquark states have been discovered in the experiment. Concerning the hexaquarks, WASA-at-COSY collaboration has reported observation of a six-quark candidate $ d^*(2380) $  with $ J^P=3^+ $ \cite{WASA-at-COSY:2014lmt}, starting a wide research on the properties of hexaquarks and dibaryons as interesting objects: a hypothetical SU(3) flavor-singlet, highly symmetric, deeply bound neutral  particle termed the scalar hexaquark $ S = uuddss $ has been introduced as a potential  candidate for dark matter  \cite{Azizi:2019xla}.   Although some resonances have been introduced as  potential condidates for the hybrids and glueballs, there are no discovered particles with high confidence level  in these categories yet. It is time to investigate the hybrid states and glueballs with a fresh eye, though they have been searched for for  a long time. Hybrid states were predicted in 1976 \cite{Jaffe:1975fd}. There are  some states with  quantum numbers  $J^{PC}={0^{--},~0^{+-},~1^{-+},~2^{-+}}$, which cannot be explained  by $q\bar{q}$ picture and are considered as potential hybrid meson candidates \cite{ParticleDataGroup:2022pth}. Among them are  $\pi_1(1400)$~\cite{IHEP-Brussels-LosAlamos-AnnecyLAPP:1988iqi}, $\pi_1(1600)$~\cite{E852:1998mbq}, $\pi_1(2015)$~\cite{E852:2004gpn}, and $\eta_1(1855)$~\cite{BESIII:2022riz} with exotic quantum numbers $J^{PC} = 1^{-+}$ evidenced  by some experiments. They are possible single-gluon hybrid candidates of a quark-antiquark pair together with a valence gluon.  Designed to search for the hybrid mesons as its primary goal,  the GlueX experiment at Jefferson Lab is expected to  give crucial insights  into the existence and structure of the exotic hybrid mesons.

Investigation of the light and heavy hybrid mesons is of great importance not only for determination of their nature and quark-gluon organization but for gaining useful information about the nonperturbative nature of QCD. Light hybrid states, which are the subject of the present study, have been intensively investigated in the framework of different theoretical methods such as lattice QCD~\cite{Dudek:2011bn,Dudek:2013yja}, the Schwinger-Dyson formalism~\cite{Burden:1996nh,Burden:2002ps,Hilger:2015ora},the flux tube model~\cite{Close:1994hc,Barnes:1995hc}, the MIT bag model~\cite{Barnes:1982zs,Chanowitz:1982qj} and QCD Laplace sum-rules(LSRs)~\cite{Balitsky:1982ps,Govaerts:1984bk,Latorre:1984kc,Govaerts:1985fx,Balitsky:1986hf,Braun:1985ah,Latorre:1985tg,Huang:1998zj,Chetyrkin:2000tj,Jin:2000ek,Jin:2002rw,Guo:2007uz,Narison:2009vj,Zhang:2013rya,Huang:2014hya,Huang:2016upt,Huang:2017pzh}. In particular, Ref.~\cite{Govaerts:1985fx} contains a comprehensive LSR analysis of the light hybrids for $J=0$ and $1$ with all the possible combinations for the parity and charge quantum numbers. Analyses show that  the $0^{++}$, $0^{--}$, $1^{++}$, and $1^{--}$ states are mainly stable, while  $0^{+-}$, $0^{-+}$, $1^{+-}$, and $1^{-+}$ quantum numbers lead to unstable and controversial results. Expected to be the lightest hybrid mesons, the ones with $1^{-+}$ have been the subject of much additional study. In the QCD sum rules, predictions on the $1^{-+}$ light hybrid mesons  are inconsistent among different works. For instance, I. I. Balitsky \textit{et al}'s prediction of  mass for $1^{-+}$ state is in the range,  1.0--1.3\,GeV \cite{Balitsky:1982ps,Balitsky:1986hf}, J . I. Latorre \textit{et al} predicted the related mass to be around 2.1\,GeV \cite{Latorre:1985tg} and the obtained mass  for  $1^{-+}$ state is around 2.5\,GeV in Refs. \cite{Govaerts:1984bk,Govaerts:1984hc,Govaerts:1986pp}. The mass of $1^{-+}$ was reanalyzed by including  the quark-gluon  condensates up to  8 dimensions  \cite{Huang:2017pzh} and it was found  that the mass value increases to be in the range 1.72--2.60\,GeV. The obtained mass range did not favor the $\pi_1(1400)$ and the $\pi_1(1600)$ to be pure hybrid states and suggests the $\pi_1(2015)$, observed by E852, to have much of a hybrid constituent.  The masses of the light hybrid mesons with $J^{PC}=1^{-+}$ quantum numbers  have also  been calculated using different theoretical methods other than QCD sum rules~\cite{Llanes-Estrada:2000lxh,Bellantuono:2014lra,Hedditch:2005zf,McNeile:2006bz}. 

Concerning other quantum numbers, the masses of the $0^{++}$ and $0^{-+}$ light hybrid mesons were calculated  using the QCD sum rule method by considering two kinds of the interpolated currents with the same quantum numbers. While the approximately equal mass was predicted for the $0^{-+}$ hybrid states from the two different currents,  different masses were obtained for the $0^{++}$ hybrid states from the two considered different currents. The masses of light (nonstrange and strange) quarkonium hybrid mesons with $J^{PC}=0^{+-}$ were investigated using Gaussian sum rules accompanied  by the Hölder inequality \cite{Ho:2018cat}. The obtained mass predictions of the nonstrange and strange  states were 2.60\,GeV and 3.57\,GeV, respectively. Unfortunately, the predictions for the masses of the light hybrid mesons with various spin and parity are inconsistent with each other. Thus, further theoretical studies for the light hybrid states with all possible quantum numbers are necessary.

In this article, inspired by this situation, we will compute the mass and current coupling of the light quarkonium hybrid mesons with different  quantum numbers. We will utilize the Borel QCD sum rule method  to carry out the calculations. This method is recognized as a powerful and predictive nonperturbative approach in the field of hadron physics~\cite{Shifman:1978bx, Shifman:1978by, Reinders:1984sr, Narison:1989aq}. It has demonstrated good success not only in analyzing the properties of conventional hadrons but also in examining the exotic particles  (see for instance Refs.  \cite{Wang:2018ejf,Wu:2018xdi,Voloshin:2018vym,Cao:2018vmv,Agaev:2020zad,Agaev:2022pis,Agaev:2023wua,Barsbay:2022gtu, Agaev:2021vur,Wang:2021itn, Azizi:2023iym, Wang:2020rdh,Wang:2018waa,Wang:2019iaa,Wang:2019hyc}). The results obtained through this approach have effectively confirmed the existing experimental data.
We construct different currents to interpolate the light quarkonium hybrid mesons with various possible spin-parities. The aim is to explore the spectroscopic characteristics of various  types of light hybrid mesons.  Note that, the scalar and vector states; and  pseudoscalar and axial-vector states, couple simultaneously  to the same currents. We isolate the contributions of different states by choosing appropriate Lorentz structures entering the calculations.

 In sec. \ref{sec:Mass}, the   sum rules for the mass and current coupling of the light strange and nonstrange  hybrid mesons of different spin-parities are derived. In sec. \ref{sec:Numeric},  the obtained sum rules  are numerically analyzed.  The last section  is  dedicated to the  summary of the calculations and conclusions. We move the expressions for the contributions of different perturbative and nonperturbative operators to the Appendix.

\section{Formalism}  \label{sec:Mass}

The sum rules for the mass and current coupling of the light hybrid mesons can be extracted  from analysis of the following  two-point correlation function:
\begin{align}
  \Pi _{\mu \nu }(q)=&i\int d^{4}xe^{iqx}\langle 0|\mathcal{T}\{J_{\mu
    }(x)J_{\nu }^{\dagger }(0)\}|0\rangle \nonumber \\
  =&\frac{q_{\mu}q_{\nu}}{q^2}\Pi_{S(PS)}(q^2) 
   + \left(\frac{q_{\mu}q_{\nu}}{q^2}-g_{\mu\nu}\right)\Pi_{V(AV)}(q^2),  \label{eq:CorrF1}
\end{align}

where $J_{\mu }(x)$ is the  current representing  the  vector ($V  $)  and  axial-vector ($AV  $) light hybrid states coupling to the scalar ($ S $) and pseudoscalar ($ PS) $ states as well. Possible interpolating currents with different valence quark-gluon contents and quantum numbers to be considered in the present study are given in Table\ \ref{tab:Current}.
\begin{table}[tbp]
\caption{The currents of the light hybrid states.}
\label{tab:Current}  
\begin{tabular}{|c|c|} 
\hline\hline
Currents & $J^{PC}$ \\  \hline\hline
$J_{\mu }^{1}=g_s \overline{q}_{a}\gamma
_{\theta}\gamma _{5} \frac{\lambda _{ab}^{n}}{2} \tilde{G}_{\mu\theta}^{n}q_{b}$ & $0^{+-},1^{--}$ \\ [2mm] \hline 
$J_{\mu }^{2}=g_s \overline{q}_{a}\gamma
_{\theta}\gamma _{5} \frac{\lambda _{ab}^{n}}{2} G_{\mu\theta}^{n}q_{b}$ & $0^{--},1^{+-}$ 
\\ [2mm]  \hline
$J_{\mu }^{3}=g_s\overline{q}_{a}\gamma
_{\theta} \frac{\lambda _{ab}^{n}}{2} \tilde{G}_{\mu\theta}^{n}q_{b}$ & $0^{-+},1^{++}$     \\ [2mm]  \hline
$J_{\mu }^{4}=g_s \overline{q}_{a}(x)\gamma
_{\theta} \frac{\lambda _{ab}^{n}}{2} G_{\mu\theta}^{n}q_{b}$ & $0^{++},1^{-+}$ \\ [2mm] \hline\hline
\end{tabular}%
\end{table}
Here   $g_{s}$ is the strong coupling constant, $a,b= 1,2,3$ are color indices, $\lambda^n$ with $n= 1,2, \cdots, 8$ are the Gell-Mann matrices,  $\tilde{G}_{\mu\theta}^{n}(x) = \epsilon_{\mu\theta\alpha\beta} G_{\alpha \beta}^{n}(x)/ 2$ is the dual field strength of $G_{\mu\theta}^{n}(x)$, and $q=u, d, s$ are the light quark fields. In Table\ \ref{tab:Current} one can see the states with different quantum numbers that  couple to each current, simultaneously. This situation causes some pollution that should be removed to calculate the physical quantities of the desired states.

To obtain the sum rules for the mass and current coupling of different states we need to relate these hadronic quantities to the fundamental QCD parameters like the quark masses, quark-gluon condensates of different nonperturbative mass dimensions, strong coupling constants and some auxiliary parameters entering the calculations at different stages based on the standard prescriptions of the method. Therefore, we need to calculate the aforesaid correlation function in two hadronic and QCD languages. Matching the coefficients of different Lorentz structures from the both representations will give us the aiming sum rules.  In technique language, the hadronic representation at time-like region is found by inserting complete sets of hadronic states with the same quantum properties as the interpolating currents between the two  creating and annihilating currents in coordinate space. As an example, for the current  $J_{\mu}^{1}$  coupling simultaneously to the  scalar $ 0^{+-} $ and vector $ 1^{--} $ states, after performing the four integrals over four-$ x $, we get

\begin{eqnarray}
&&\Pi _{\mu \nu }^{\mathrm{Phys}}(q)=\frac{\langle 0|J_{\mu }^{1}|H_{S}(q)\rangle
\langle H_{S}(q)|J_{\nu }^{1\dagger }|0\rangle }{m_{H_{S}}^{2}-q^{2}}   \notag \\
&&+\frac{\langle 0|J_{\mu }^{1}|H_{V}(q)\rangle
\langle H_{V}(q)|J_{\nu }^{1\dagger }|0\rangle }{m_{H_{V}}^{2}-q^{2}}+\ldots,  \notag \\
&&  \label{eq:PhysSide}
\end{eqnarray}%
where $\Pi _{\mu \nu }^{\mathrm{Phys}}(q) $ stands for the physical or hadronic side; and $m_{H_{S}}$ and $m_{H_{V}}$ are the masses of the light  $ S $ and $V  $ hybrid states, respectively.  To proceed, we need to define the following matrix elements:
\begin{equation}
\langle 0|J_{\mu }|H_{S}(q)\rangle =q_{\mu}f_{H_{S}}\,
\label{eq:Mel1}
\end{equation}%
and 
\begin{equation}
\langle 0|J_{\mu }|H_{V}(q)\rangle =m_{H_{V}}f_{H_{V}}\varepsilon _{\mu },
\label{eq:Mel1}
\end{equation}%
where $\varepsilon _{\mu }$ is the polarization vector of the $  V$ state; and $f_{H_{S}}$ and $f_{H_{V}}$ represent the corresponding current couplings. Up to here, the function $\Pi _{\mu \nu }^{\mathrm{Phys}}(q)$ takes the form,
\begin{align}
\Pi _{\mu \nu }^{\mathrm{Phys}}(q)=&\frac{f_{H_{S}}^{2}}{m_{H_{S}}^{2}-q^{2}}q_{\mu }q_{\nu }+\frac{m_{H_{V}}^{2}f_{H_{V}}^{2}}{m_{H_{V}}^{2}-q^{2}} \nonumber \\
&\times\left( -g_{\mu \nu }+\frac{q_{\mu }q_{\nu }}{q^{2}}\right) +\cdots. 
\label{eq:Phys}
\end{align}%
As it is clear, the function  $\Pi _{\mu \nu }^{\mathrm{Phys}}(q)$  contains the ${S}$ and ${V}$ hybrid contributions, simultaneously. To isolate the ${S}$ contribution alone, it is enough to multiply Eq.\ (\ref{eq:Phys})  by $q_{\mu }q_{\nu }/{q^{2}}$ which results in,
\begin{equation}
\frac{q_{\mu }q_{\nu }}{q^{2}}\Pi _{\mu \nu }^{\mathrm{Phys}}(q)=-f_{H_{S}}^{2}+\frac{m_{H_{S}}^{2}f_{H_{S}}^{2}}{m_{H_{S}}^{2}-q^{2}} +\cdots.
\label{eq:Phys1}
\end{equation}%
The contribution of  vector state is found by  choosing the structure $g_{\mu \nu }$ which is free of the  ${S}$ hybrid state and contains only the vector contribution:
\begin{equation}
\Pi _{\mu \nu }^{\prime\mathrm{Phys}}(q)=\frac{m_{H_{V}}^{2}f_{H_{V}}^{2}}{m_{H_{V}}^{2}-q^{2}}\times\left( -g_{\mu \nu }\right) +\cdots. 
\label{eq:Physp}
\end{equation}%
To suppress the  contributions of the higher states and continuum at each channel, we apply  Borel transformation with respect to $q^{2}$.  We find, 
\begin{equation}
\mathcal{B}_{q^{2}}\Pi _{\mu \nu }^{\prime\mathrm{Phys}%
}(q)=m_{H_{V}}^{2}f_{H_{V}}^{2}e^{-m_{H_{V}}^{2}/M^{2}}\left( -g_{\mu \nu }\right) +\cdots,  \label{eq:CorBor1}
\end{equation}
and
\begin{equation}
\mathcal{B}_{q^{2}}\frac{q_{\mu }q_{\nu }}{q^{2}}\Pi _{\mu \nu }^{\mathrm{Phys}}(q)=m_{H_{S}}^{2}f_{H_{S}}^{2}e^{-m_{H_{S}}^{2}/M^{2}}\ +\cdots .  \label{eq:CorBor2}
\end{equation}
for the final representations of the physical contributions of the $  V$ and $  S$ states in Borel scheme, respectively. In these equations $ M^2 $ is the Borel parameter to be fixed in next section.

On the QCD side, the correlation function is obtained in deep Euclidean space-like region, $ q^2\rightarrow -\infty $, where the time ordering operator of two currents are expanded in terms of different perturbative and nonperturbative contributions using  operator product expansion (OPE). We take into account the nonperturbative contributions up to ten dimensions. In fact, to do this, we insert the explicit forms of the currents into the correlation function and perform all the possible contractions among the quark fields in vacuum. The result is obtained in terms of the participating quarks propagators and vacuum expectation value of the two valence gluons. As an example, for the current $J_{\mu}^{1}$ the function $\Pi _{\mu \nu }^{\mathrm{OPE}}(q)$ takes the form
\begin{eqnarray}
&&\Pi _{\mu \nu }^{\mathrm{OPE}}(q)=\frac{g_s^{2}\epsilon_{\mu\theta\alpha\beta}\epsilon_{\nu\delta\alpha^{\prime }\beta^{\prime }}}{4}i\int d^{4}xe^{iqx}  \notag \\
&&\times \langle 0|G^n_{\alpha \beta}(x) G^m_{\alpha^{\prime }\beta^{\prime }}(0)|0 \rangle \mathrm t^{n}[a,b] \mathrm t^{m}[a^{\prime },b^{\prime }] \notag \\
&&\times \mathrm{Tr}\left[S_{q}^{a^{\prime }a}(-x)\gamma _{\theta }\gamma _{5}S_{q}^{bb^{\prime }}(x)\gamma _{5}\gamma _{\delta }\right],  \notag \\
&&  \label{eq:OPE1}
\end{eqnarray}%
where $t^n={\lambda^n}/2$ and  $S_{q}^{ab}(x)$ is the light-quark propagator, whose expression in coordinate space is  given by
\begin{eqnarray}
&&S_{q}^{ab}(x)=i\frac{\slashed x\delta _{ab}}{2\pi ^{2}x^{4}}-\frac{%
m_{q}\delta _{ab}}{4\pi ^{2}x^{2}}-\frac{\langle \overline{q}q\rangle }{12}%
\left( 1-i\frac{m_{q}}{4}\slashed x\right) \delta _{ab}  \notag \\
&&-\frac{x^{2}}{192}\langle \overline{q}g_{s}\sigma Gq\rangle \left( 1-i%
\frac{m_{q}}{6}\slashed x\right) \delta _{ab}-\frac{\slashed xx^{2}g_{s}^{2}%
}{7776}\langle \overline{q}q\rangle ^{2}\delta _{ab}  \notag \\
&&-\frac{ig_{s}G_{ab}^{\mu \nu }}{32\pi ^{2}x^{2}}\left[ \slashed x\sigma
_{\mu \nu }+\sigma _{\mu \nu }\slashed x\right] -\frac{x^{4}\langle
\overline{q}q\rangle \langle g_{s}^{2}G^{2}\rangle }{27648}\delta
_{ab}+\cdots.  \notag \\
&&  \label{eq:qProp}
\end{eqnarray}%
When replacing this propagator into Eq.\ (\ref{eq:OPE1}), besides the different perturbative and nonperturbative contributions of different dimensions resulting from the multiplications of two propagators, there appears another term from the multiplications of  $ -\frac{ig_{s}G_{ab}^{\mu \nu }}{32\pi ^{2}x^{2}}\left[ \slashed x\sigma_{\mu \nu }+\sigma _{\mu \nu }\slashed x\right] $ from both the propagators.  This term makes  another two-gluon condensate  in the presence of vacuum.  We apply the following  short-hand notations: 
\begin{eqnarray}
G_{ab}^{\alpha \beta } &=&G_{A}^{\alpha \beta
}\lambda _{ab}^{A}/2,\,\,~~G^{2}=G_{\alpha \beta }^{A}G_{\alpha \beta }^{A},  
\end{eqnarray}%
in the calculations.

Now, we proceed to discuss the matrix element  $ \langle 0 |G^n_{\alpha \beta}(x)G^m_{\alpha' \beta'}(0)|0 \rangle $ in Eqs.\ (\ref{eq:OPE1}). It is considered by two different ways:  In the first approach, it is considered as the full gluon propagator in coordinate space between two points $ 0 $ and $ x $, whose expression is given by
\begin{eqnarray}
\label{eq:Gprop}
&&\langle 0 |G^n_{\alpha \beta}(x)G^m_{\alpha' \beta'}(0)||0 \rangle =
\frac{\delta^{mn}}{2 \pi^2 x^4} [g_{\beta \beta'}(g_{\alpha \alpha'}-\frac{4 x_{\alpha} x_{\alpha'}}{x^2}) \nonumber \\
&& +(\beta, \beta') \leftrightarrow (\alpha, \alpha')
 -\beta \leftrightarrow \alpha -\beta' \leftrightarrow \alpha'].
\end{eqnarray}
This is multiplied to the two light quark propagators and all the calculations are done.   This contribution  is  equal to the diagrams at which the two quarks and the   valence-gluon  all are  full propagators.  In the second approach,  the matrix element $ \langle 0 |G^n_{\alpha \beta}(x)G^m_{\alpha' \beta'}(0)|0 \rangle $ is considered as the two-gluon condensate. To this end the gluon field at point   $ x $  is  expanded  around $x=0$  and the first term is taken into account. Hence, one has,
\begin{eqnarray}
\label{eq:Gcond}
&&\langle 0 |G^n_{\alpha \beta}(0)G^m_{\alpha' \beta'}(0)|0 \rangle =
\frac{\langle G^2\rangle }{96}\delta^{mn} [g_{\alpha \alpha'} g_{\beta \beta'} \notag \\
&&-g_{\alpha \beta'} g_{\alpha'\beta }].
\end{eqnarray}
The contribution coming from this approach is equivalent to   the diagrams  with  full quark propagators and the gluon interacting with the QCD vacuum.
By  using  Eq.\ (\ref{eq:Gprop}) or (\ref{eq:Gcond}) into  Eqs.\ (\ref{eq:OPE1}), we  make use of the following relation:
\begin{eqnarray}
\label{eq:tntn}
&&\mathrm t^{n}[a,b] \mathrm t^{n}[a^{\prime },b^{\prime }]=\frac{1}{2}\left(\delta^{ab'}\delta^{a'b}-\frac{1}{3}\delta^{ab}\delta^{a'b'}\right).
\end{eqnarray}
After putting all the elements explained above together, we find expressions in OPE side in terms of fundamental QCD parameters like the quark masses, quark and gluon condensates, strong coupling constant etc. in coordinate space. We apply Fourier transformation to transfer the calculations to the momentum space. To this end, we make use of the  formula
\begin{eqnarray}
\label{ }
\frac{1}{(x^2)^m}&=&\int \frac{d^Dk }{(2\pi)^D}e^{-ik \cdot x}i(-1)^{m+1}2^{D-2m}\pi^{D/2} \nonumber \\
&\times& \frac{\Gamma[D/2-m]}{\Gamma[m]}\Big(-\frac{1}{k^2}\Big)^{D/2-m},
\end{eqnarray}
then perform integral over four-$ x $. To perform the integrals over other four-parameters, we use the Feynman parametrization and the formula,
\begin{equation}
\label{ }
\int d^4 \ell\frac{(\ell^2)^m}{(\ell^2+\Delta)^n}=\frac{i\pi^2 (-1)^{m-n} \Gamma[m+2]\Gamma[n-m-2]}{\Gamma[2]\Gamma[n] (-\Delta)^{n-m-2}},
\end{equation} 
where, $ \Delta $ depends on the parameters of the problem except $ \ell $. Finally, we use 
\begin{equation}
\label{ }
\Gamma\Big[\frac{D}{2}-n\Big]\Big(-\frac{1}{\Delta}\Big)^{D/2-n}=\frac{(-1)^{n-1}}{(n-2)!}(-\Delta)^{n-2}ln[-\Delta].
\end{equation} 
to find the imaginary parts of the resultant expressions for the perturbative and nonperturbative contributions  up to six dimensions, which gives the spectral densities in dispersion representation. Note that,  for the contributions equal to or greater than the six dimensions, we directly find the contributions of the nonperturbative operators based on the standard prescriptions of the method. The next step is applying the Borel transformation to the OPE side the same as the physical side.  The final procedure, is to apply continuum subtraction to further suppress the contributions of higher states and continuum. When performing this, a threshold $ s_0 $ is set to use the quark-hadron duality assumption. The continuum threshod depends on the energy of the first excited state and will be fixed in next section as well. 
As a result, for the OPE side we get
\begin{equation}
\Pi (M^{2},s_{0})=\int_{4m_{q}^{2}}^{s_{0}}ds\rho ^{\mathrm{OPE}%
}(s)e^{-s/M^{2}}+\Gamma (M^{2}),  \label{eq:InvAmp}
\end{equation}%
where $\rho ^{\mathrm{OPE}}(s)=\frac{1}{\pi}Im[ \Pi (q^2)]$ is the two-point spectral density calculated up to dimension six nonperturbative contributions. The
 invariant amplitude $\Gamma (M^{2})$ contains nonperturbative contributions calculated directly from  the OPE side and for the nonperturbative contributions from six to ten mass dimensions. Note that the six dimension appears both in the $\rho ^{\mathrm{OPE}} (s) $ and $ \Gamma (M^{2}) $.   The explicit expressions of these  functions are presented  in the  Appendix,  as samples, for the strange hybrid meson with the quantum numbers $J^{PC}=1^{--}$.

The last task in this section is to choose appropriate Lorentz structures from both sides and match the invariant coefficients. 
We shall say that, for our case,  these structures are $ g_{\mu \nu }$ and $ I$, which typify contributions of $ V (AV) $ and $ S (PC) $ particles, respectively. For the mass $m_{H}$ and current coupling $f_{H}$ of the  hybrid  states  associated to the current  $J_{\mu}^{1}$, we get
\begin{equation}
m_{H}^{2}=\frac{d/dx[\Pi (M^{2},s_{0})]}{\Pi (M^{2},s_{0})},  \label{eq:Mass}
\end{equation}
and
\begin{equation}
f_{H}^{2}=\frac{e^{m_{H}^{2}/M^{2}}}{m_{H}^{2}}\Pi (M^{2},s_{0}),  \label{eq:Coupling}
\end{equation}%
where $ x= -1/M^{2}$.  Similar computations are done for the currents $J_{\mu}^{2}$, $J_{\mu}^{3}$ and $J_{\mu}^{4}$ to get sum rules for the physical quantities of the corresponding hybrid states.

\section{Numerical Results}
\label{sec:Numeric}
In this section, we perform numerical analysis to get the values of the masses and current couplings. The quark masses are taken from PDG, while for the quark-gluon condensates, we use: 
\begin{eqnarray}
&&\langle \overline{q}q\rangle =-(0.24\pm 0.01)^{3}~\mathrm{GeV}^{3},\
\notag \\
&&\langle \overline{s}s\rangle =0.8~\langle \overline{q}q\rangle ,\
\notag \\
&&\langle \overline{q}g_{s}\sigma Gq\rangle =m_{0}^{2}\langle \overline{q}%
q\rangle ,\ m_{0}^{2}=(0.8\pm 0.1)~\mathrm{GeV}^{2},\   \notag \\
&&\langle \frac{\alpha _{s}G^{2}}{\pi }\rangle =(0.012\pm 0.004)~\mathrm{GeV}%
^{4}.
 \label{eq:Parameters}
\end{eqnarray}%
The higher dimensional operators up to ten are obtained in terms of the condensates presented above (see the Appendix).

Besides the above-mentioned input parameters, the obtained sum rules in the previous section depend on two auxiliary parameters: The Borel parameter  $M^{2}$ and the continuum threshold $s_{0}$. Their working windows are found based on the standard requirements of the method. These are the relatively weak dependence of the physical quantities, dominance of pole contribution at each channel and convergence of the OPE. For the convergence of OPE series we need that the perturbative part exceeds the total nonperturbative contribution and the higher the dimension of the nonperturbative operator the lower its contribution.  In technique language, we define the following formula for the pole contribution ($  PC$):
\begin{equation}
\mathrm{PC}=\frac{\Pi(M^{2},\ s_{0})}{\Pi(M^{2},\ \infty )} . \label{eq:Cond1}
\end{equation}
The condition $ PC\geq 0.5 $ sets the higher limit of the Borel parameter $M^{2}$. Its lower limit  is fixed by the convergence of the OPE series. For this, we employ the condition 
\begin{equation}
R(M_{\mathrm{min}}^{2})=\frac{\Pi^{\mathrm{DimN}}(M_{\mathrm{%
min}}^{2},\ s_{0})}{\Pi(M_{\mathrm{min}}^{2},\ s_{0})}\leq0.05,
\label{eq:Cond2}
\end{equation}
where $\Pi ^{\mathrm{DimN}}(M_{\mathrm{min}}^{2},s_{0})$ with  $\mathrm{DimN=Dim(8+9+10)}$ denotes the contributions of the last three nonperturbative operators.  The continuum threshold $ s_0 $ depends on the energy of the first excited states at each channel and it is fixed such that the above conditions are satisfied and we witness the relatively weak dependence of the mass and current coupling with respect to its variations in the working window. The residual dependence arising from the determination of the working intervals of the $M^{2}$ and $ s_0 $ appear as the uncertainties of the results. The uncertainties coming from the auxiliary parameters constitute the main errors in the presented values.  The intervals for the $ M^2 $ and  the values of $ PC $ at lower and higher values of $ M^2 $ for all the considered channels at average values of the continuum threshold are given  in tables\ \ref{ss_PCresults_table} and \ref{qq_PCresults_table} for the strange and nonstrange hybrid mesons, respectively. The presented values for $ PC $ in these tables  are acceptable for the exotic states. We will present the exact intervals of the continuum thresholds for each channel in the following tables.  
As an example, the perturbative and different nonperturbative contributions (in units of  $ GeV^8 $) for  the strange hybrid meson of the  $J^{PC}=1^{--}$ quantum numbers  as functions of $M^2$  and at  average $s_0$ are depicted in Fig.\ \ref{fig:Contribution} . As it is seen, the OPE convergence condition is nicely satisfied.

\begin{widetext}

\begin{table}[!ht]
\centering
\caption{Values of $ PC $  and   intervals of  $ M^2 $   for the ground state light strange hybrid states  at average value of the continuum threshold. }
\label{ss_PCresults_table}
\begin{tabular}{cccc}
  $J^{PC}$ & $s_{0}(average)\ (GeV^2)$ & $[M_{\mathrm{min}}^{2},M_{\mathrm{max}}^{2}]\ (GeV^2)$
  & $\mathrm{PC}$ \\ 
\hline
  $0^{++}$ & $19.0$ & $[5.0,6.0]$ & $ [0.27,0.22] $\\
  $0^{+-}$ & $17.0$ & $[4.5,5.5]$ & $ [0.64,0.48] $\\
  $0^{--}$ & $19.0$ & $[5.5,6.5]$ & $ [0.22,0.17] $\\
  $0^{-+}$ & $18.0$ & $[4.5,5.5]$ & $ [0.68,0.52] $\\
  $1^{++}$ & $10.0$ & $[4.5,5.5]$ & $ [0.08,0.05] $\\
  $1^{+-}$ & $9.0$ & $[3.0,4.0]$ & $ [0.48,0.27] $\\
  $1^{--}$ & $9.0$ & $[2.5,3.5]$ & $ [0.46,0.25] $\\
  $1^{-+}$ & $9.0$ & $[2.5,3.5]$ & $ [0.62,0.36] $ 
\end{tabular}
\end{table}

\begin{table}[!ht]
\centering
\caption{Values of $ PC $  and   intervals of  $ M^2 $   for  the ground state light nonstrange hybrid states  at average value of the continuum threshold.  }
\label{qq_PCresults_table}
\begin{tabular}{cccc}
  $J^{PC}$ & $s_{0}(average)\ (GeV^2)$ & $[M_{\mathrm{min}}^{2},M_{\mathrm{max}}^{2}]\ (GeV^2)$
  & $\mathrm{PC}$ \\ 
\hline
  $0^{++}$ & $19.0$ & $[5.5,6.5]$ & $ [0.23,0.18] $\\
  $0^{+-}$ & $17.0$ & $[4.5,5.5]$ & $ [0.65,0.48] $\\
  $0^{--}$ & $19.0$ & $[5.5,6.5]$ & $ [0.23,0.18] $\\
  $0^{-+}$ & $17.0$ & $[4.5,5.5]$ & $ [0.65,0.48] $\\
  $1^{++}$ & $10.0$ & $[4.5,5.5]$ & $ [0.08,0.05] $\\
  $1^{+-}$ & $9.0$ & $[2.5,3.5]$ & $ [0.63,0.36] $\\
  $1^{--}$ & $9.0$ & $[2.5,3.5]$ & $ [0.48,0.26] $\\
  $1^{-+}$ & $9.0$ & $[2.5,3.5]$ & $ [0.63,0.36] $
\end{tabular}
\end{table}

\begin{figure}[h!]
\begin{center}
\includegraphics[totalheight=6cm,width=8cm]{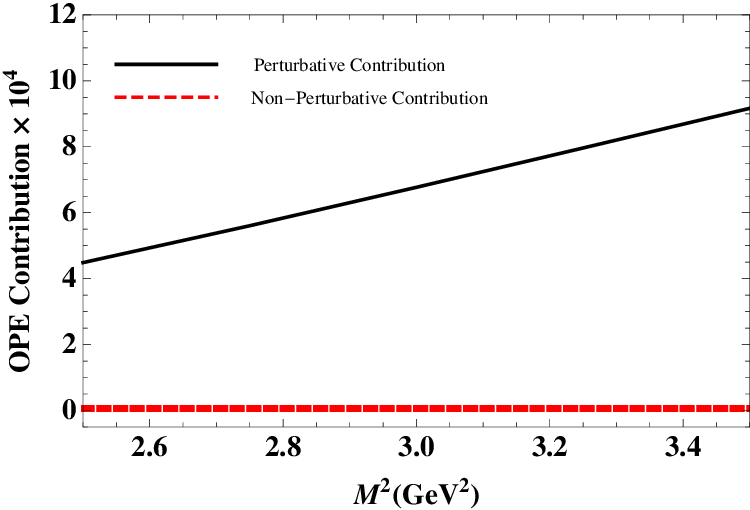}\,\, %
\includegraphics[totalheight=6cm,width=8cm]{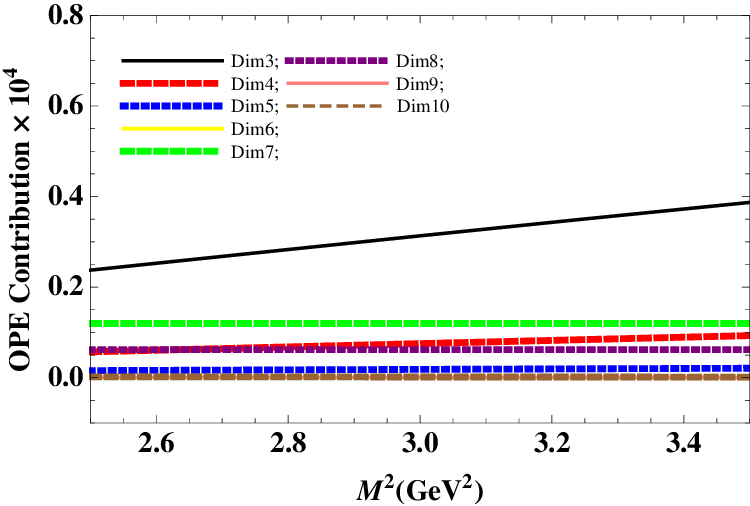}
\end{center}
\caption{The perturbative and different nonperturbative contributions (in units of  $ GeV^8 $) for  the strange hybrid meson of the  $J^{PC}=1^{--}$ quantum numbers  as functions of $M^2$  and at  average $s_0$. }
\label{fig:Contribution}
\end{figure}

\end{widetext}

Having calculated the working intervals of  the auxiliary parameters, we depict the variations of the mass  $m_{H}$ and current coupling $f_{H}$  as functions of the Borel parameter and continuum threshold  in Figs.\ \ref{fig:Mass} and  \ref{fig:Decay constant}. We observe that the  physical quantities demonstrate good stability  with respect to the variations  of the    parameters $M^2$  and $s_0$  in their working intervals. This is the case for other members and quantum numbers as well. 
\begin{widetext}

\begin{figure}[h!]
\begin{center}
\includegraphics[totalheight=6cm,width=8cm]{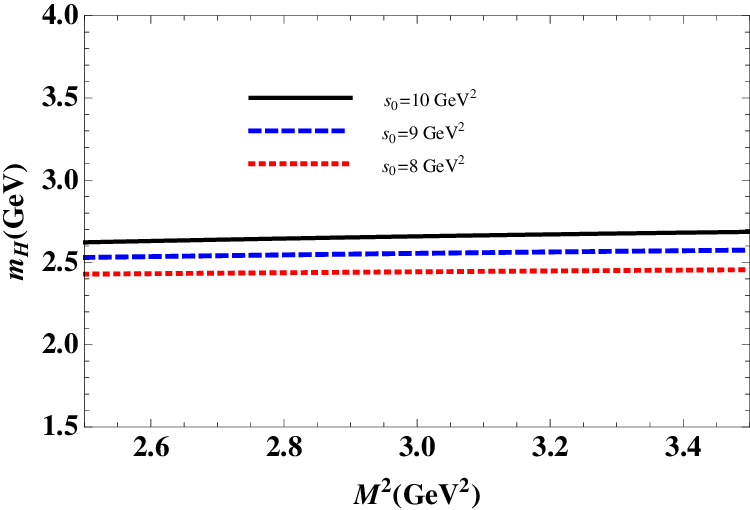}\,\, %
\includegraphics[totalheight=6cm,width=8cm]{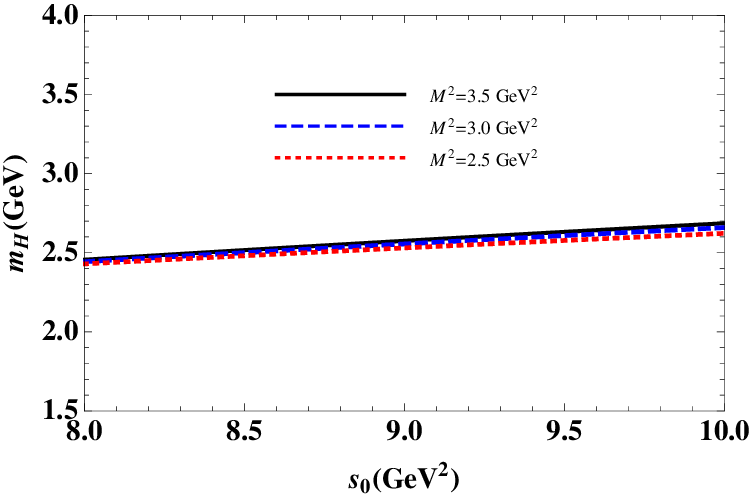}
\end{center}
\caption{ The mass of the strange hybrid meson with  $J^{PC}=1^{--}$ as functions of  $M^2$  and  $s_0$.}
\label{fig:Mass}
\end{figure}

\begin{figure}[h!]
\begin{center}
\includegraphics[totalheight=6cm,width=8cm]{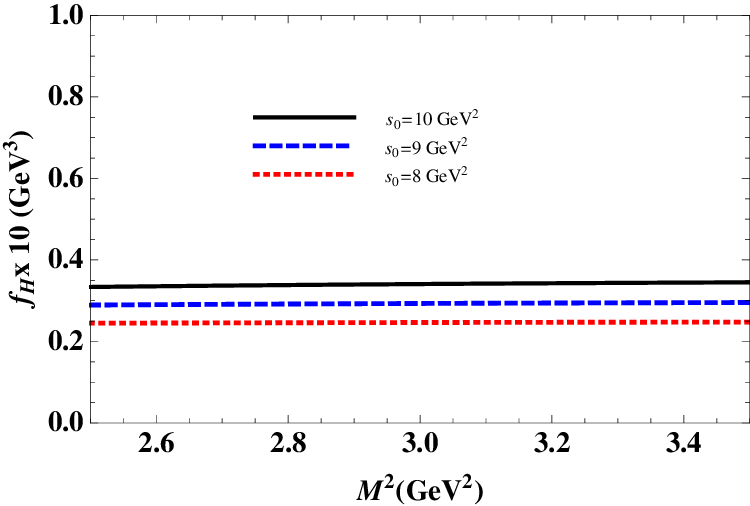}\,\, %
\includegraphics[totalheight=6cm,width=8cm]{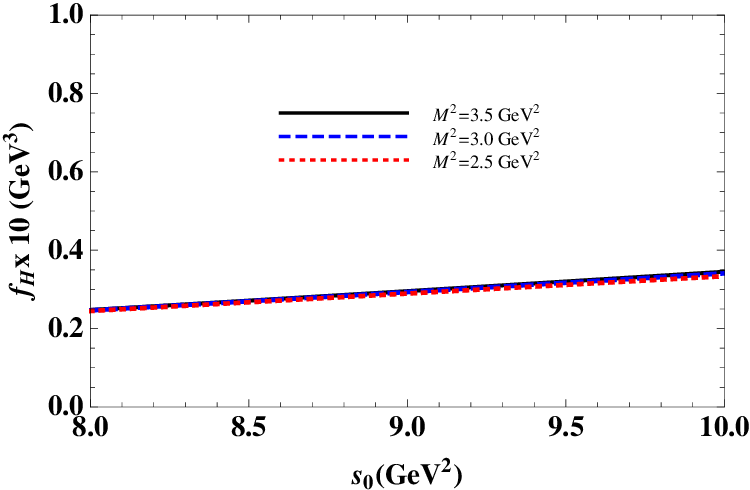}
\end{center}
\caption{ The current coupling of  the strange hybrid meson with  $J^{PC}=1^{--}$ as functions of  $M^2$  and  $s_0$.}
\label{fig:Decay constant}
\end{figure}

\begin{table}[!ht]
\centering
\caption{Values of the mass and current coupling for different quantum numbers of the ground state strange hybrid mesons. }
\label{ss_results_table}
\begin{tabular}{ccccc}
  $J^{PC}$ & $M^2 \pm \delta M^2\ (GeV^2)$
  & $s_0 \pm \delta s_0\ (GeV^2)$  & $m_H \pm \delta m_{H}\ (GeV)$ & $(f_{H}\pm \delta f_{H})\times10\ (GeV^3)$\\ 
\hline
  $0^{++}$ & $5.5\pm 0.5$ & $19.0\pm 1.0$  & $ 4.58^{+0.29}_{-0.17}$ & $ 0.63^{+0.11}_{-0.11}$\\
  $0^{+-}$ & $5.0\pm 0.5$ & $17.0\pm 1.0$  & $ 3.14^{+0.14}_{-0.13}$ & $ 0.65^{+0.05}_{-0.04}$\\
  $0^{--}$ & $6.0\pm 0.5$ & $19.0\pm 1.0$  & $ 4.64^{+0.29}_{-0.17}$ & $ 0.59^{+0.09}_{-0.10}$\\
  $0^{-+}$ & $5.0\pm 0.5$ & $18.0\pm 1.0$  & $ 3.20^{+0.14}_{-0.13}$ & $ 0.69^{+0.05}_{-0.05}$\\
  $1^{++}$ & $5.0\pm 0.5$ & $10.0\pm 1.0$  & $ 3.22^{+0.13}_{-0.07}$ & $ 0.26^{+0.06}_{-0.06}$\\
  $1^{+-}$ & $3.5\pm 0.5$ & $9.0\pm 1.0$  & $ 2.35^{+0.17}_{-0.16}$ & $ 0.38^{+0.05}_{-0.05}$\\
  $1^{--}$ & $3.0\pm 0.5$ & $9.0\pm 1.0$  & $ 2.55^{+0.14}_{-0.12}$ & $ 0.29^{+0.05}_{-0.04}$\\
  $1^{-+}$ & $3.0\pm 0.5$ & $9.0\pm 1.0$  & $ 2.31^{+0.18}_{-0.16}$ & $ 0.38^{+0.05}_{-0.04}$
\end{tabular}
\end{table}

\begin{table}[!ht]
\centering
\caption{Values of the mass and current coupling for different quantum numbers of the ground state nonstrange hybrid mesons. }
\label{qq_results_table}
\begin{tabular}{ccccc}
    $J^{PC}$ & $M^2 \pm \delta M^2\ (GeV^2)$
  & $s_0 \pm \delta s_0\ (GeV^2)$  & $m_H \pm \delta m_{H}\ (GeV)$ & $(f_{H}\pm \delta f_{H})\times10\ (GeV^3)$\\ 
\hline
  $0^{++}$ & $6.0\pm 0.5$ & $19.0\pm 1.0$  & $ 4.53^{+0.21}_{-0.13}$ & $ 0.58^{+0.08}_{-0.09}$\\
  $0^{+-}$ & $5.0\pm 0.5$ & $17.0\pm 1.0$  & $ 3.13^{+0.14}_{-0.13}$ & $ 0.65^{+0.05}_{-0.04}$\\
  $0^{--}$ & $6.0\pm 0.5$ & $19.0\pm 1.0$  & $ 4.52^{+0.21}_{-0.13}$ & $ 0.58^{+0.08}_{-0.08}$\\
  $0^{-+}$ & $5.0\pm 0.5$ & $17.0\pm 1.0$  & $ 3.14^{+0.14}_{-0.13}$ & $ 0.65^{+0.05}_{-0.04}$\\
  $1^{++}$ & $5.0\pm 0.5$ & $10.0\pm 1.0$  & $ 3.19^{+0.11}_{-0.07}$ & $ 0.26^{+0.06}_{-0.06}$\\
  $1^{+-}$ & $3.0\pm 0.5$ & $9.0\pm 1.0$  & $ 2.30^{+0.18}_{-0.17}$ & $ 0.38^{+0.06}_{-0.03}$\\
  $1^{--}$ & $3.0\pm 0.5$ & $9.0\pm 1.0$  & $ 2.51^{+0.15}_{-0.14}$ & $ 0.30^{+0.05}_{-0.04}$\\
  $1^{-+}$ & $3.0\pm 0.5$ & $9.0\pm 1.0$  & $ 2.30^{+0.18}_{-0.17}$ & $ 0.38^{+0.05}_{-0.04}$
\end{tabular}
\end{table}

\begin{table}[tbp]
	\centering
	\caption{The masses of the ground state strange hybrid mesons in $\mathrm{GeV}$ and comparison of the results with other predictions.}
	\label{ssbargtab1}
	\begin{tabular}{|c|c|c|c|c|c|c|}
		
		\hline   $J^{PC}$& This work & Ref. \cite{Govaerts:1985fx} &Ref. \cite{Balitsky:1986hf} & Ref. \cite{Latorre:1985tg} &  Ref. \cite{Guo:2007uz} &    Ref. \cite{Ho:2018cat}  \\ \hline\hline
		$	0^{++}$& $4.58^{+0.29}_{-0.17}$                         &$3.5$&$-$&	$-$&$-$&-	  \\  
		$	0^{+-}$&$3.14^{+0.14}_{-0.13}$                                   &$3.4$&$-$&	$-$&$-$&$3.57$	 \\ 
		$	0^{--}$&$4.64^{+0.29}_{-0.17}$                 &$3.7$&	$-$&$3.8  $ &$-$&-  \\
		$	0^{-+}$&$3.20^{+0.14}_{-0.13}$                              & $3.1$&$-$ &	$-$&$-$&-	 \\ 
		$	1^{++}$&$3.22^{+0.13}_{-0.07}$   & $2.8$&$-$&$-$ & $-$	&-	\\
		$	1^{+-}$	&$2.35^{+0.17}_{-0.16}$              &$2.8$&$-$&$-$ & $-$	&- \\ 
		$	1^{--}$&$2.55^{+0.14}_{-0.12}$          &$2.9$&$-$&	$-$ & $2.54-2.62$&-	   \\ 
		$	1^{-+}$&$2.31^{+0.18}_{-0.16}$       &$2.5$&$\sim1.75$&	$\sim1.6-2.1$ & $-$&-	\\ 
		
		\hline\hline
	\end{tabular}%
\end{table}   
\begin{table}[tbp]
	\centering
	\caption{The masses of the ground state nonstrange hybrid mesons in $\mathrm{GeV}$ and comparison of the results with other predictions.}
	\label{qqbargtab1}
	\begin{tabular}{|c|c|c|c|c|c|c|c|c|c|c|c|}
		
		\hline   $J^{PC}$& This work & Ref. \cite{Latorre:1984kc} &Ref. \cite{Balitsky:1986hf} & Ref. \cite{Braun:1985ah} &  Ref. \cite{Huang:1998zj} & Ref. \cite{Jin:2000ek}  & Ref. \cite{Guo:2007uz} & Ref. \cite{Narison:2009vj} & Ref. \cite{Zhang:2013rya} & Ref. \cite{Huang:2014hya} & Ref. \cite{Ho:2018cat}  \\ \hline\hline
		$	0^{++}$& $4.53^{+0.21}_{-0.13}$ &$-$&$-$&	$-$&$2.35-3.4$&$1.75-2.00$&$-$&$-$&$>4$&$-$&-	  \\  
		$	0^{+-}$&$3.13^{+0.14}_{-0.13}$  &$-$&$-$&$\sim2.1-2.5$&$-$&$-$&$-$&$-$&$-$& & $2.60$	 \\ 
		$	0^{--}$&$4.52^{+0.21}_{-0.13}$  &$3.1  \pm0.2$&	$-$&$- $ &$-$&$-$&$-$&$-$&$-$&$-$&-  \\
		$	0^{-+}$&$3.14^{+0.14}_{-0.13}$  & $-$&$-$ &$-$&$2.3$&$-$&$-$&$-$&	$-$&$-$&-	 \\ 
		$	1^{++}$&$3.19^{+0.11}_{-0.07}$  & $-$&$-$&$-$&$-$&$-$&$-$&$-$&$-$ & $-$	&-	\\
		$	1^{+-}$	&$2.30^{+0.18}_{-0.17}$ &$-$&$-$&$-$ &$-$&$-$&$-$&$-$&$-$& $-$	&-\\ 
		$	1^{--}$&$2.51^{+0.15}_{-0.14}$  &$-$&$-$&	$-$ & $-$&$-$&$2.33-2.43$&$-$&$-$&$-$&-	   \\ 
		$	1^{-+}$&$2.30^{+0.18}_{-0.17}$  &$1.7  \pm0.1$&$\sim1.5$&	$-$ & $-$&$1.55$&$-$&$1.86-2.2$&$1.71  \pm0.22$&$1.72-2.60$&-	\\ 
		
		\hline\hline
	\end{tabular}%
\end{table}

\end{widetext}

Now, we proceed to present the values of the masses and current couplings of  the light hybrid mesons with the quantum numbers $	0^{++}$,  $	0^{+-}$, $	0^{--}$, $	0^{-+}$, $	1^{++}$, $	1^{+-}$,  $	1^{--}$ and $	1^{-+}$ for the strange and nonstrange members in tables\ \ref{ss_results_table}-\ref{qq_results_table}, respectively.  For completeness, we also depict the intervals of the Borel parameter $ M^2 $ and continuum threshold $ s_0 $ for all the channels in these tables. The uncertainties present in the values of the masses and current couplings are due to the uncertainties of the auxiliary parameters and errors of other input values. The order of uncertainties in the values of the masses are lower compared to those of the current coupling constants. This is due the fact that the mass is ratio of two sum rules, whose uncertainties partly  kill each other, while the current coupling is found from one sum rule and receives relatively more errors.  The masses of the ground state light hybrid mesons for the strange and nonstrange members are available from different sources, as well. However, they have been presented either without uncertainties or are not available for all the members and quantum numbers. In this view, our results calculated with a higher accuracy can be a useful guide for the related experimental groups at different hadron colliders in their search for these yet unseen interesting particles. The comparison of our results for the strange and nonstrange members with other existing predictions in the literature are made in tables \ \ref{ssbargtab1} and  \ref{qqbargtab1}, respectively. As it is seen from the tables, there is a consistency among the presented results in some channels within the uncertainties, but we see some serious differences of the presented values of different sources for some other channels. The previous studies done by QCD sum rules method are up to five, six, or eight mass dimensions of nonperturbative operators and do not include the uncertainties. This is the case for other frameworks as well. An exact comparison can be made, when the predictions from other studies/approaches  are available for all the strange and nonstrange light hybrid mesons with different quantum numbers with their uncertainties.

At the end of this section, we would like to study the finite width effect on the parameters of the states under study.  This causes the following modification in the hadronic side of the mass sum rule compared to the  zero-width single-pole approximation \cite{Wang:2015nwa,Sundu:2018nxt}: 
\begin{equation}
\frac{1}{m^{2}_H-q^{2}}\rightarrow \frac{1}{m^{2}_H-q^{2}-i\sqrt{q^{2}}\Gamma_H}, \label{eq:Modification}
\end{equation}
where $ \Gamma_H $ is the width of the hybrid state. In the Borel scheme,  the corresponding sum  rule becomes:
\begin{eqnarray}
\label{finitewidth}
&&\frac{2}{\pi}m_{H}^{3}f_{H}^{2} \Gamma_{H}\times \nonumber\\&&\int^{\infty}_0\frac{e^{-s/M^2}}{[s-m_{H}^{2} ]^{2}+m_{H}^{2}\Gamma_{H}^{2}}=\Pi (M^2,s_0) .\nonumber\\
\end{eqnarray}%
To find the new mass, current coupling and width as three unknowns,  we need to have two more equations, which are found by  successive applications of the operator $ \frac{d}{d(-\frac{1}{M^{2}})} $  to both sides of the above sum rule. By simultaneous numerical  solving of the  resultant three equations,  one finds the  $ m_H $,  $ f_H $ and $\Gamma_H  $.  We present the results of these quantities  in tables \ref{ss_results_table2}  and  \ref{nss_results_table2}. From these tables, we see that the finite width effects modify the masses and current couplings, considerably. We hope that the results for the masses and widths obtained in this study  will help experimental groups to identify such quarkonium hybrid mesons.

\begin{widetext}

\begin{table}[!ht]
\centering
\caption{Values of the mass, current coupling and width for different quantum numbers of the ground state strange hybrid mesons after taking into account the  finite width effects. }
\label{ss_results_table2}
\begin{tabular}{cccc}
  $J^{PC}$   & $m_{H} \pm \delta m_{H}\ (GeV)$ & $(f_{H}\pm \delta f_{H})\times10\ (GeV^3)$ & $ \Gamma_{H}  \pm \delta \Gamma_{H}\ (MeV)$\\ 
\hline
  $0^{++}$  & $ 4.06^{+0.26}_{-0.15}$ & $ 0.32^{+0.05}_{-0.05}$& $ 216^{+14}_{-8}$\\
  $0^{+-}$  & $ 3.55^{+0.16}_{-0.15}$ & $ 0.40^{+0.03}_{-0.02}$& $ 405^{+18}_{-17}$\\
  $0^{--}$  & $ 4.04^{+0.25}_{-0.15}$ & $ 0.48^{+0.07}_{-0.08}$& $ 252^{+16}_{-9}$\\
  $0^{-+}$ & $ 3.63^{+0.16}_{-0.15}$ & $ 0.42^{+0.03}_{-0.03}$& $ 403^{+18}_{-16}$\\
  $1^{++}$ & $ 3.54^{+0.14}_{-0.08}$ & $ 0.19^{+0.04}_{-0.04}$& $ 298^{+12}_{-6}$\\
  $1^{+-}$  & $ 2.64^{+0.19}_{-0.18}$ & $ 0.24^{+0.03}_{-0.03}$& $ 298^{+22}_{-20}$\\
  $1^{--}$ & $ 2.65^{+0.14}_{-0.12}$ & $ 0.19^{+0.03}_{-0.02}$& $ 114^{+6}_{-5}$\\
  $1^{-+}$ & $ 2.60^{+0.20}_{-0.18}$ & $ 0.23^{+0.03}_{-0.02}$& $ 304^{+24}_{-21}$
\end{tabular}
\end{table}

\begin{table}[!ht]
\centering
\caption{Values of the mass, current coupling and width for different quantum numbers of the ground state nonstrange hybrid mesons after taking into account the  finite width effects. }
\label{nss_results_table2}
\begin{tabular}{cccc}
  $J^{PC}$   & $m_{H} \pm \delta m_{H}\ (GeV)$ & $(f_{H}\pm \delta f_{H})\times10\ (GeV^3)$ & $ \Gamma_{H}  \pm \delta \Gamma_{H}\ (MeV)$\\ 
\hline
  $0^{++}$  & $ 3.98^{+0.18}_{-0.11}$ & $ 0.37^{+0.05}_{-0.06}$& $ 385^{+18}_{-11}$\\
  $0^{+-}$  & $ 3.54^{+0.16}_{-0.15}$ & $ 0.40^{+0.03}_{-0.02}$& $ 423^{+19}_{-18}$\\
  $0^{--}$  & $ 4.01^{+0.19}_{-0.12}$ & $ 0.42^{+0.06}_{-0.06}$& $ 297^{+14}_{-8}$\\
  $0^{-+}$ & $ 3.55^{+0.16}_{-0.15}$ & $ 0.40^{+0.03}_{-0.02}$& $ 415^{+18}_{-17}$\\
  $1^{++}$ & $ 3.52^{+0.12}_{-0.07}$ & $ 0.19^{+0.04}_{-0.04}$& $ 285^{+10}_{-6}$\\
  $1^{+-}$  & $ 2.60^{+0.20}_{-0.19}$ & $ 0.23^{+0.04}_{-0.02}$& $ 317^{+25}_{-23}$\\
  $1^{--}$ & $ 2.83^{+0.17}_{-0.16}$ & $ 0.17^{+0.03}_{-0.02}$& $ 194^{+12}_{-11}$\\
  $1^{-+}$ & $ 2.60^{+0.20}_{-0.19}$ & $ 0.23^{+0.03}_{-0.02}$& $ 317^{+25}_{-23}$
\end{tabular}
\end{table}
\end{widetext}

\section{Summary and Concluding Notes}
\label{sec:Conc}
The QCD as quantum field theory of strong interaction and the quark model as the most powerful model for the categorizing the particles made of quarks and gluons allow the existence of exotic states out of the usual mesons and baryons/antibaryons. The existence of exotic states of different quark-gluon configurations was predicted by Jaffe in the mid-1970s. However, the first observation  of exotic states was made in 2003 by discovery of  the tetraquark $ X(3872) $  by Belle collaboration. After that many tetraquark and pentaquark states of charmed-anticharmed content were observed by different collaborations. By the progresses made in the experiments on the identification  of doubly heavy baryons,   the discovery of the first doubly heavy charmed tetraquak  $T^+_{cc}(3875)$ was reported by the LHCb collaboration in 2021 that was another  revolutionary discovery of the exotic states. Considering these developments, it is not  surprising to see other categories of the exotic states like hybrid mesons, glueballs, etc.

Inspired by the experimental and theoretical progress on the exotic states, we investigated the light strange and nonstrange hybrid mesons with quantum numbers $	0^{++}$,  $	0^{+-}$, $	0^{--}$, $	0^{-+}$, $	1^{++}$, $	1^{+-}$,  $	1^{--}$ and $	1^{-+}$ in the present study. We calculated the mass and  current coupling of all the members up to a high accuracy, nonperturbative operator of ten dimension.  We had the  vector-scalar and axialvecor-pseudoscalar mixing as they couple to the same currents: They are separated by the techniques mentioned in section II. We found the working intervals of the auxiliary parameters based on the standard requirements of the method to make the physical observables like mass and current coupling possibly independent of these helping parameters. We obtained the stable sum rules and presented the numerical results for the mass and current coupling of all the strange and nonstrange members of $	0^{++}$,  $	0^{+-}$, $	0^{--}$, $	0^{-+}$, $	1^{++}$, $	1^{+-}$,  $	1^{--}$ and $	1^{-+}$ quantum numbers. We also presented and compared the previously existing predictions in the literature.  We observed that the finite width effects modify the masses and current couplings of the hybrid mesons, considerably.
We hope that, our results obtained with a high accuracy will help the experimental groups in their search for the light hybrid mesons. Comparison of any future experimental data on the considered states with our predictions will shed light on the nature, quark-gluon configuration and inner structure of these interesting particles.

\section*{ACKNOWLEDGMENTS}

The work of H. S. and B. B. was supported in part by the TUBITAK via the grant  No: 123F197. K. A.  thanks  Iran National Science Foundation  (INSF)
for the partial financial support provided under the elites Grant No. 4025036. We dedicate this paper to Durmu\c{ s} Ali Demir for his important contributions to developments of  physics in the country: He left us too early.

\appendix*

\begin{widetext}

\section{ Some expressions in the OPE side of the calculations for  the strange hybrid meson of  $J^{PC}=1^{--}$}

\renewcommand{\theequation}{\Alph{section}.\arabic{equation}} \label{sec:App}

In this appendix, as examples, we present the explicit expressions of the invariant functions $\rho ^{\mathrm{OPE}} (s)  $ and $ \Gamma (M^{2}) $ for the light strange hybrid meson with quantum numbers $J^{PC}=1^{--}$. We can write,
\begin{equation}
\rho ^{\mathrm{OPE}}(s)=\rho ^{\mathrm{pert.}}(s)+\sum_{N=3}^{6}\rho ^{%
\mathrm{DimN}}(s),\ \ \Gamma (M^{2})=\sum_{N=6}^{10}\Gamma ^{\mathrm{DimN}}(M^{2}),
\label{eq:A1}
\end{equation}%
where,
\begin{equation}
\rho ^{\mathrm{DimN}}(s)=\int_{0}^{1}d%
\alpha \rho ^{\mathrm{DimN}}(s),  \label{eq:A2}
\end{equation}%
and
\begin{equation}
\Gamma ^{\mathrm{DimN}%
}(M^{2})=\int_{0}^{1}d\alpha\Gamma  ^{\mathrm{DimN}}(M^{2},\alpha ).
\label{eq:A4}
\end{equation}%

The perturbative and nonperturbative components of the spectral density are found as
\begin{eqnarray}
&&\rho ^{\mathrm{pert.}}(s)=\frac{g_{s}^{2}s^{2}(-5m_{s}^{2}+2s)}{2^{9}\cdot 3\cdot 5\pi
^{4}} ,
\end{eqnarray}%
\begin{equation}
\rho ^{\mathrm{Dim3}}(s)=\frac{g_{s}^{2}m_{s}\langle \overline{s}s\rangle s}{2^{2}\cdot 3^{2}\pi
^{2}} ,
\end{equation}%
\begin{eqnarray}
&&\rho ^{\mathrm{Dim4}}(s)=\langle \alpha _{s}G^{2}/\pi
\rangle\left[\frac{g_{s}^{2}s}{2^{6}\cdot 3^{3}\pi ^{2}}-\frac{(-9m_{s}^{2}+2s)}{2^{5}\cdot 3^{2}} \right],
\end{eqnarray}%
\begin{eqnarray}
&&\rho ^{\mathrm{Dim5}}(s )=-\frac{g_{s}^{2}m_{o}^{2}m_{s}\langle \overline{s}s\rangle}{2^{6}\cdot 3\pi ^{2}} ,
\end{eqnarray}%
and
\begin{eqnarray}
&&\rho ^{\mathrm{Dim6}}(s)=\frac{g_{s}^{4}\langle \overline{s}s\rangle ^{2}}{2^{4}\cdot 3^{4}\pi ^{2}} .
\end{eqnarray}%

Components of the function $\Gamma (M^{2})$ are:%
\begin{eqnarray}
&&\Gamma ^{\mathrm{Dim6}}(M^{2})=-\frac{g_{s}^{2}m_{s}^{2}\langle \overline{s}s\rangle ^{2}}{2^{4}\cdot 3} ,
\end{eqnarray}%

\begin{eqnarray}
&&\Gamma ^{\mathrm{Dim7}}(M^{2})=\frac{m_{s}\langle \alpha _{s}G^{2}/\pi
\rangle \langle \overline{s}s \rangle}{2^{3}\cdot 3^{2}}\left[\frac{-g_{s}^{2}}{2^{2}\cdot 3}+{7\pi ^{2}}\right] ,
\end{eqnarray}%

\begin{eqnarray}
&&\Gamma ^{\mathrm{Dim8}}(M^{2})=-\frac{g_{s}^{2}m_{o}^{2}\langle \overline{s}s \rangle ^{2}}{2^{3}\cdot 3^{2}}\left[1+\frac{m_{s}^{2}}{2M^{2}}\right]-5\frac{\langle \alpha
_{s}G^{2}/\pi \rangle ^{2} \pi ^{2}}{2^{6}\cdot 3^{3}} ,
\end{eqnarray}%

\begin{eqnarray}
&&\Gamma ^{\mathrm{Dim9}}(M^{2})=\frac{m_{s}\langle \overline{s}s \rangle}{2^{2}\cdot 3^{3}M^{2}}\left[\frac{g_{s}^{4}\langle \overline{s}s \rangle ^{2}}{3^{2}}-\frac{m_{o}^{2}\pi ^{2}\langle \alpha _{s}G^{2}/\pi
\rangle}{2^{2}}\right],
\end{eqnarray}%

\begin{eqnarray}
&&\Gamma  ^{\mathrm{Dim10}}(M^{2} )=\frac{g_{s}^{2}\langle \overline{s}s \rangle ^{2}}{2^{2}\cdot 3^{2}M^{2}}\left[-\frac{m_{o}^{4}}{2^{3}}-\langle \alpha _{s}G^{2}/\pi
\rangle\pi ^{2}(\frac{1}{3^{2}}-\dfrac{1}{3^{4}})\right].
\end{eqnarray}%

\end{widetext}

\renewcommand{\theequation}{\Alph{section}.\arabic{equation}} \label{sec:App}



\end{document}